\documentclass[aps,prl,twocolumn,groupedaddress]{revtex4}
\usepackage{graphicx}

\newcommand{\mum}{~$\mu$m}
\begin{document}


\title{Microwave inductance of thin metal strips}

\author{Katrin Steinberg}
\affiliation{1.~Physikalisches Institut, Universit\"at Stuttgart, Pfaffenwaldring 57, D-70550 Stuttgart, Germany}
\author{Marc Scheffler}
\affiliation{1.~Physikalisches Institut, Universit\"at Stuttgart, Pfaffenwaldring 57, D-70550 Stuttgart, Germany}
\author{Martin Dressel}
\affiliation{1.~Physikalisches Institut, Universit\"at Stuttgart, Pfaffenwaldring 57, D-70550 Stuttgart, Germany}

\date{\today}

\begin{abstract}
We have measured the frequency-dependent, complex impedance of thin metal strips in a broad range of microwave frequencies (45~MHz to 20~GHz). The spectra are in good agreement with theoretical predictions of an RCL model. The resistance, inductance, and capacitance, which govern the microwave response, depend on the strip width and thickness as well as on the strip and substrate materials.
While the strip resistance scales inversely with the cross section, the inductance depends on the width of the strip, but not on the thickness (in the limit of small thickness). 
\end{abstract}

\maketitle


To properly design and thoroughly analyze the inductance of electrical circuitry at high frequencies is essential for a wide range of commercial applications as well as in fundamental research. The most basic element here is a conducting wire, which might have the shape of a spiral coil\cite{yue2000,cao2003} or act as an interconnect.\cite{tunnell2008,li2009} 
The inductance of a simple geometry might be calculated easily,\cite{grover1946,terman1943} but these calculations fail for more complex structures. For  GHZ frequencies, finding the appropriate model for a realistic device can be challenging due to features like skin effect, coupling between different conductors, self inductance of thin strips, and the possible influence of a substrate. These effects become more complex if the material has a frequency-dependent conductivity. It is therefore necessary to distinguish between effects caused by geometry and by material properties. To test the validity of a possible model, one should compare it to measured data. In this work we concentrate on the simplest possible inductive structure, a straight metallic wire. This is in contrast to recent studies which address certain specific, device-relevant inductive structures.\cite{yue2000,cao2003,tunnell2008,li2009} We study a large variety of thin metallic strips, and we present microwave measurements covering an extremely broad frequency range up to 20~GHz. 

\begin{figure}
    \centering
        \includegraphics[width=8.5cm]{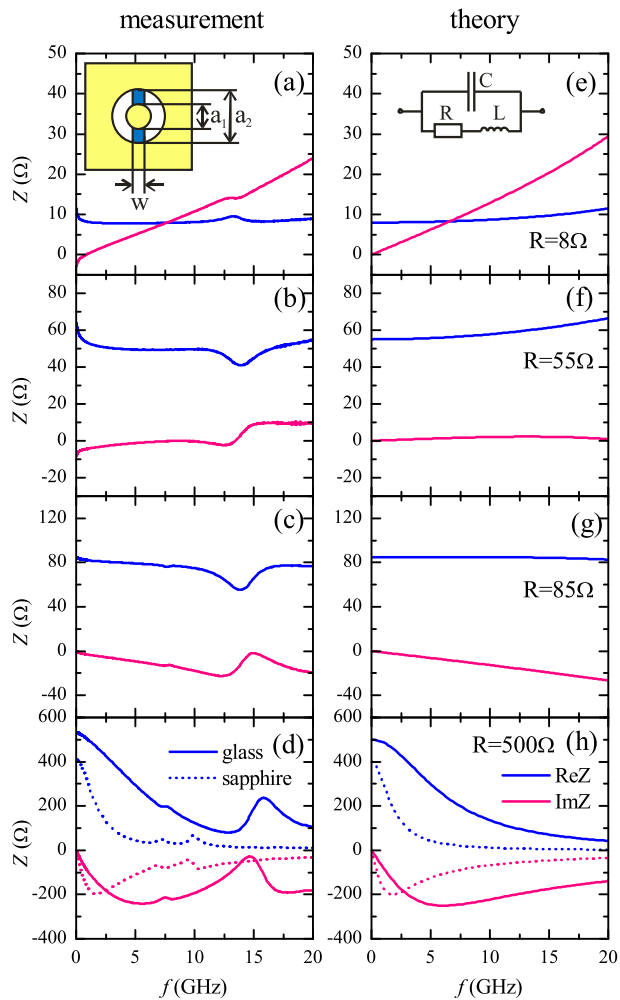}
    \caption{\label{fig:Fig1} (Color online) Real and imaginary parts of measured (left panels) and calculated (right panels) strip impedances. The measured strips have widely varying resistance $R$ due to material and thickness [(a) Pb, $t=100$~nm, $w=80$\mum; (b) Al, $t=10$~nm, $w=70$\mum; (c) NiCr, $t=60$~nm, $w=50$\mum; (d) NiCr, $t=25$~nm, $w=50$\mum{} (glass substrate), $w=60$\mum{} (sapphire substrate)]. All spectra in the right panels were calculated with the RCL model using the same values for $L=0.2~\textrm{nH}$ and $C=55~\textrm{fF}$. They differ only in the resistance [(e) $R=8\Omega$; (f) $R=55\Omega$; (g) $R=85\Omega$ ; (h) $R=500\Omega$], which leads to the different frequency dependences from inductance-dominated spectra to capacitance-dominated spectra.}
\end{figure}

Similar to leads on circuit boards or in chips, our wires are created by deposition of a thin metallic film on top of a dielectric, i.e.\ the strips have thickness $t$, width $w$, and length $l$, with $t \ll w < l$. We will describe this \lq device\rq{} with the lumped elements shown in the inset of Fig.~\ref{fig:Fig1}(e): the wire has a resistance $R$ in series with its inductance $L$, and in parallel the capacitance $C$ due to the dielectric substrate has to be taken into account. The complex impedance of this circuit as a function of angular frequency $\omega = 2 \pi f$ is:
\begin{equation}\label{eq:Z}
    Z_{\text{Sample}} = \frac{R+i (\omega(L-R^2C) - \omega^3L^2C)}{1-\omega^2(R^2C^2-2LC)+\omega^4 L^2C^2}
\end{equation}
We will consider frequencies as high as 20~GHz, but even then the skin depth will be much larger than $t$, and therefore $R$ and $L$ will only depend on the wire geometry and material (and not on frequency). For such a strip, the geometric or Faraday inductance is given by \cite{yue2000,Greenhouse1974}
\begin{equation}\label{eq:greenhouse}
    L[\textrm{nH}]=0.2 l[\textrm{mm}]\left[ \ln\left(\frac{2l}{w+t}\right)+0.5+\frac{w+t}{3l}\right]
\end{equation}
where $L$ is the inductance in nH, and the length $l$, the width $w$, and the thickness $t$ are in mm.

The dimension of the strip that affects the inductance most strongly is the length $l$. The width $w$ has a much weaker influence, and the strip thickness can be neglected completely for our limit ($t \ll w$). This is in contrast to the resistance $R = \rho l /(w t)$, which is inversely proportional to the strip cross section $w t$ and depends on the material properties via its resistivity $\rho$. The frequency dependence of the impedance $Z_{\text{Sample}}$, Eq.~(\ref{eq:greenhouse}), can differ drastically, depending on the values of $L$, $C$, and $R$. For certain combinations of $L$ and $C$, this also holds if these two quantities are kept constant and only $R$ is varied. In Fig.~\ref{fig:Fig1} we demonstrate this by combining experimental and theoretical data.

In the experiment, we compare numerous samples. Each consists of a metallic strip, prepared by thermal evaporation of nichrome (NiCr), lead (Pb), or aluminum (Al) and deposition onto a glass or sapphire substrate through a shadow mask. The width and thickness of the different strips vary over substantial ranges. The microwave measurements are performed at room temperature with a Corbino spectrometer\cite{scheffler2005a} that is suitable for this particular geometry,\cite{scheffler2007} as shown in the inset of Fig.~\ref{fig:Fig1}(a): concentric gold contacts in Corbino geometry (inner diameter $a_1=0.8~\mathrm{mm}$, outer  diameter $a_2=1.75~\mathrm{mm}$) are deposited on top of the strip. We then measure the complex reflection coefficient, which directly reveals the sample impedance.\cite{scheffler2005a,stutzman2000} We have performed a full three standard calibration using as references an Al plate (short), a NiCr thin film (load), and a Teflon cylinder (open). Due to the particular geometry, the sample impedance is that of two strips [with length $l=(a_2-a_1)/2$, width $w$, and thickness $t$] in parallel.

In Figs.\ \ref{fig:Fig1}(a)-(d) we present experimental impedance spectra spanning the very broad frequency range $f$=~45~MHz to $f$=~20~GHz. $a_1$=1.75~mm and $a_2$=0.8~mm are kept constant for all measurements. For these data, our choices of $t$, $w$, and strip material only weakly affect $L$ and $C$ of the sample, but drastically change $R$. Therefore, we compare the experimental data to theoretical impedance spectra, Figs.\ \ref{fig:Fig1}(e)-(h), where we keep $L=0.2$~nH and $C=55$~fF constant, but vary $R$. In general, we can describe the experimental data very closely with the simple RCL model, except for an impedance increase toward the lowest frequencies [in Figs.~\ref{fig:Fig1}(a), (b), due to capacitive contributions to the contact impedance]\cite{steinberg2008} and a resonance feature around 12-15~GHz for glass substrates and 8-10~GHz for sapphire substrates (due to a volume mode governed by the dielectric substrate).\cite{scheffler2010,kitano2008}

Tuning $R$ over a large range, we find rather different frequency dependences, depending on the ratio $R/\sqrt{2L/C}$:
For $R \ll \sqrt{2L/C}$ [Figs.\ \ref{fig:Fig1}(a), (e)], Re($Z(f)$) is almost constant whereas Im($Z(f)$) roughly increases linearly with frequency. For $R$ slightly smaller than $\sqrt{2L/C}$ [Figs.\ \ref{fig:Fig1}(b), (f)], Im($Z(f)$) almost vanishes whereas Re($Z(f)$)$\approx$$R$ is almost constant. For slightly larger $R$, now exceeding $\sqrt{2L/C}$, Re($Z(f)$) develops a downturn toward higher frequencies, and Im($Z(f)$) is negative, decreasing linearly with frequency. For the extreme case $R \gg \sqrt{2L/C}$, Re($Z(f)$) starts at Re($Z$($f$=0))=$R$ and decreases strongly with increasing frequency, roughly as 1/$f^2$, whereas Im($Z(f)$)$<$0 first decreases linearly with $-f$ and then approaches zero from below as -1/$f$; the minimum occurs for $f$=1/(2$\pi RC$). For this last case,\cite{scheffler2007} we additionally present data for a slightly thicker NiCr strip deposited on a substrate of sapphire (instead of glass). Due to the larger dielectric constant of sapphire, the capacitance $C$ between inner and outer Corbino contact increases and the characteristic frequency dependence in Figs.~\ref{fig:Fig1}(d), (h) moves to lower frequencies.

\begin{figure}[htbp]
    \centering
        \includegraphics[width=8.5cm]{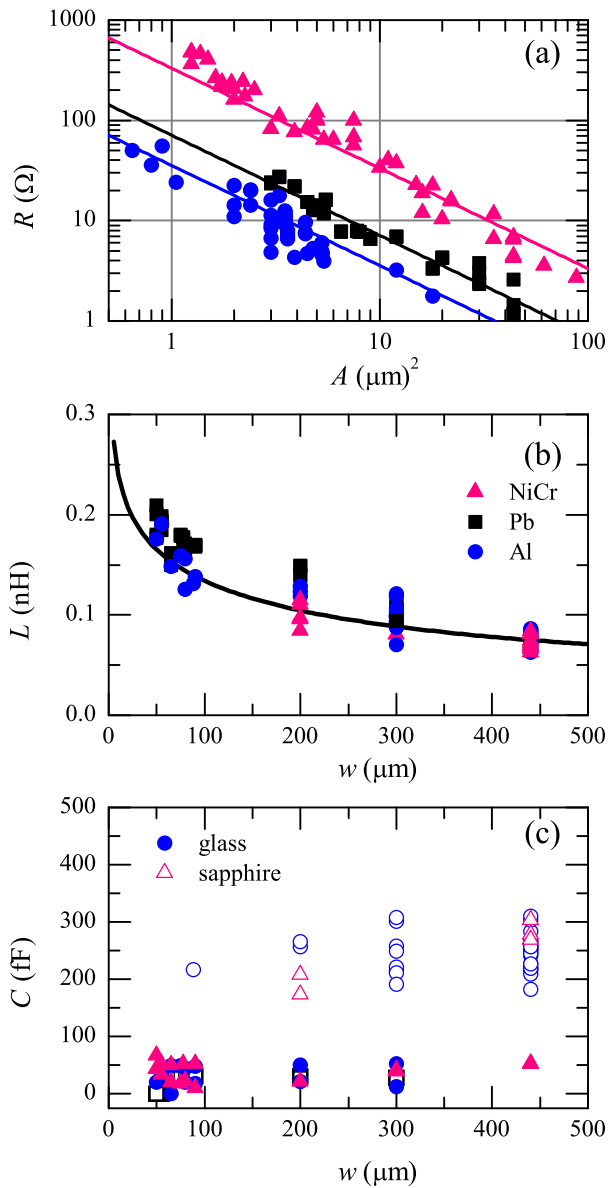}
    \caption{\label{fig:Fig2} (Color online) Resistance $R$, inductance $L$, and capacitance $C$ obtained from RCL fits to the impedance spectra of all strip samples. $R$ is plotted versus cross section $A=w t$ of the strips while $L$ and $C$ are plotted against the strip width \textit{w}. Full line in Fig.~\ref{fig:Fig2}(b) is the prediction by Eq.~(\ref{eq:greenhouse}). The capacitance [Fig.~\ref{fig:Fig2}(c)] differs for the cases of glass (filled symbols) and sapphire (empty symbols) substrates.}
\end{figure}

The real and imaginary parts of the measured impedance spectra were simultaneously fitted with Eq.~(\ref{eq:Z}) to obtain  $R$, $L$, and $C$ for different strip geometries.\cite{CommentFit} As evident from Fig.~\ref{fig:Fig1}, the measured spectra and the model spectra agree quite well; thus $R$, $L$, and $C$ can be determined from these fits. The fitting error of these parameters vary depending on the particular regime of the impedance spectrum: while $R$ can always be determined reliably, $C$ has a large fitting error except for the large-$R$ case [Fig.~\ref{fig:Fig1}(d)], where the pronounced frequency dependence is governed by $C$. Similarly, $L$ can precisely be determined for small $R$ [Fig.~\ref{fig:Fig1}(a)], but in addition also for those cases, where $R$ and $C$ both are large [like the sapphire case in Fig.~\ref{fig:Fig1}(d)].

The results of these fits are shown in Fig. \ref{fig:Fig2}. The resistance $R=\rho l / A$ of a strip is expected to depend on the dc resistivity $\rho$ of the strip material and the strip cross section $A= w t$. This is exactly what we see in Fig.~\ref{fig:Fig2}(a): for each material (NiCr, Pb, and Al), $R$ depends inversely on $A$ (as indicated by the straight line, which is a fit with $\rho$ as fitting parameter: $\rho_{\rm NiCr}=128~\mu\Omega$cm, $\rho_{\rm Al}=14~\mu\Omega $cm, $\rho_{\rm Pb}= 31~\mu \Omega$cm. The imprecise determination of the film thickness of the aluminum samples leads to data the data scattering. 
More relevant for our interest is the inductance [Fig.~\ref{fig:Fig2}(b)] that we obtain from the fits: according to Eq.~(\ref{eq:greenhouse}) we expect that for our samples ($l$ constant; $w \ll t$) the inductance will only depend on $w$, and this is what we find. Independent of strip material and thickness, the width dependence of $L$ follows Eq.~(\ref{eq:greenhouse}) as indicated by the full line in Fig.~\ref{fig:Fig2}(b).
For the capacitance as the third fitting parameter, on the other hand, the dependence on $w$ should only be weak,\cite{scheffler2007} but it should depend strongly on the dielectric constant of the substrate material. This is seen in Fig.~\ref{fig:Fig2}(c); the devices on sapphire substrates (open symbols) have a much higher capacitance than those on glass substrates (closed symbols), and within each substrate material there is, within the data scattering, no width dependence. The scattering of the data is  mostly due to the inaccurate determination of $C$ for inductance dominated samples.

We have investigated the microwave impedance of thin metallic strips and we could successfully describe their frequency dependence with an RCL model. In particular, we have shown how the geometric inductance contributes to the overall impedance and how it depends on the strip width. This is not only relevant for the design of microwave devices, but also for fundamental research. For example, for several materials of current interest, such as superconductors,\cite{steinberg2008,booth1996} interacting electron systems,\cite{scheffler2010,scheffler2005b} or two-dimensional electron gases,\cite{burke2000} the intrinsic conductivity is frequency dependent even for frequencies as low as a few GHz. When these materials are studied using microwaves, the imaginary part of the complex conductivity shows up as an inductive contribution. In this case, one has to know the geometric inductance, as presented in this Paper, before one can determine the intrinsic inductance stemming from the material.

We thank Gabriele Untereiner for the sample fabrication and Ren\'e Ramsperger and Serife Kilic for some of the microwave measurements. We acknowledge financial support by the DFG and SFB/TRR21.

\end{document}